\documentclass[12pt,a4paper]{article}
\usepackage[a4paper,margin=1.5in]{geometry}
\usepackage[dvips]{graphicx}
\usepackage{color}
\usepackage{cite}
\usepackage{hyperref}
\usepackage[cmex10]{amsmath}
\usepackage{amsthm}
\usepackage{amssymb}
\usepackage{algorithmic}
\usepackage{url}
\usepackage{algorithm}
\usepackage{subfigure}
\usepackage{wrapfig}
\usepackage[normalem]{ulem}
\newcommand{\stkout}[1]{\ifmmode\text{\sout{\ensuremath{#1}}}\else\sout{#1}\fi}
{}
{}
{}

\usepackage{url}
\usepackage{float}

\begin{document}

\title{Non-normality Can Facilitate Pulsing in Biomolecular Circuits}

\author{Abhilash Patel, Shaunak Sen\\
Electrical Engineering\\
Indian Institute of Technology Delhi\\
Hauz Khas, New Delhi 110016, India.\\
Email: shaunak.sen@ee.iitd.ac.in
}
\date{}
{\color{blue}\textit{This paper is a preprint of a paper accepted by IET Systems Biology and is subject to Institution of Engineering and Technology Copyright. When the final version is published, the copy of record will be available at the IET Digital Library.}
}
{\let\newpage\relax\maketitle}

\begin{abstract}
Non-normality can underlie pulse dynamics in many engineering contexts.
However, its role in pulses generated in biomolecular contexts is generally unclear.
Here, we address this issue using the mathematical tools of linear algebra and systems theory on simple computational models of biomolecular circuits.
We find that non-normality is present in standard models of feedforward loops.
We used a generalized framework and pseudospectrum analysis to identify non-normality in larger biomolecular circuit models, finding that it correlates well with pulsing dynamics.
Finally, we illustrate how these methods can be used to provide analytical support to numerical screens for pulsing dynamics as well as provide guidelines for design.
\end{abstract}

\maketitle

     
\section{Introduction}
The output of a system is said to exhibit a pulse when there is a transient growth followed by a decay to equilibrium (Fig. 1a).
This is in contrast to a trajectory exhibiting a direct decay to equilibrium.
The dynamics in larger systems may need to be described by multiple decaying trajectories, or their combinations, corresponding to the principal degrees of freedom.
Normally, the combinations of multiple decaying trajectories itself decays to the equilibrium.
However, there are significant set of combinations, such as a difference of two decaying trajectories, that can pulse.
These are at the core of the mathematical idea of non-normality, which is associated with the transition from laminar flow to turbulence~\cite{reshotko01}.
Non-normality may be present in multiple contexts in science and engineering.
\begin{figure}[!tbh]
\centering
\includegraphics[width=3.8in]{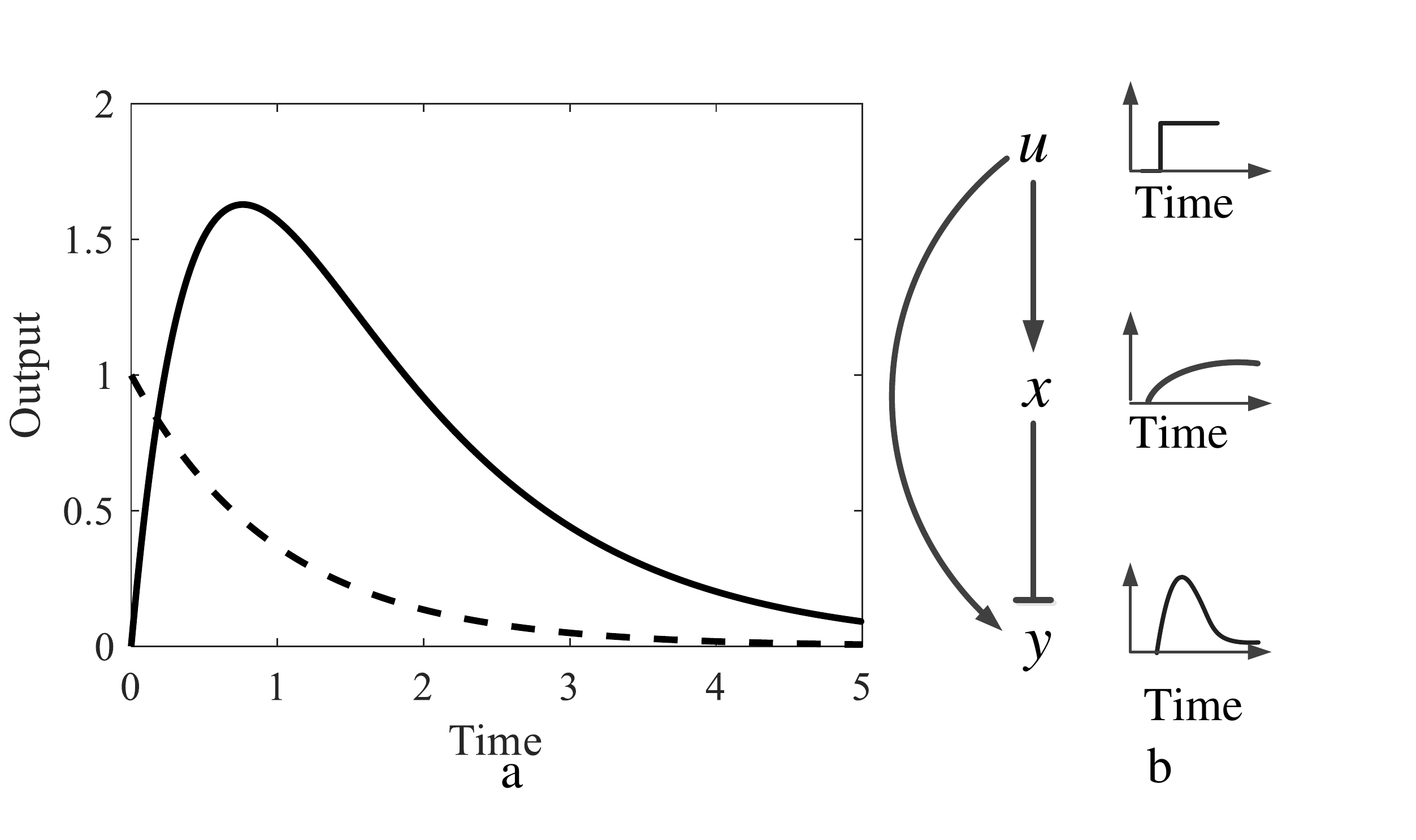}
\caption{a) Black solid line represents a pulse that grows and decays to equilibrium. Black dashed line represents a decay to equilibrium. b) Illustration of an incoherent feedforward loop. Graphs indicate typical trajectories.}
\label{fig:0}
\end{figure}

Recent advances in high resolution temporal imaging have highlighted the functional importance of pulse behaviour in biomolecular systems as well as their underlying dynamical mechanisms~\cite{levine13}.
One example of pulse behaviour is in the transient differentiation process of competence in the bacteria \textit{Bacillus subtilis}~\cite{locke2011}.
It has been shown that the underlying circuit is composed of both positive and negative feedback that can generate excitability, a parameter regime where some small amplitude random fluctuations can generate large amplitude pulses in the output.
Similar mechanisms are believed to underlie neuron dynamics~\cite{morozova2016dopamine}.
Further, such pulsed behaviour is also seen in systems exhibiting perfect adaptation, such as bacterial chemotaxis~\cite{alon2006book}.
In such systems, the output exhibits a transient deviation from a fixed value on the application of a step change in stimulus.
This transient deviation often takes the form of a pulse and can be viewed as a temporal derivative to a step change in the input.
An important mechanism underlying this adaptation is integral feedback~\cite{yi00}, where the difference between the output and its desired value is integrated in a feedback loop so as to ensure that the output converges to its desired value.
Finally, another biomolecular circuit that can generate pulses is an incoherent feedforward loop~\cite{kaplan08}, an overrepresented motif in genetic networks~\cite{mangan03}.
In a simple realization (Fig. 1b), a step input \textit{u} activates the expression of two proteins \textit{x} and \textit{y}.
The concentrations of both proteins start to increase.
Here, \textit{x} is also a transcriptional repressor of \textit{y} and, due to the repressing action of \textit{x} on \textit{y}, the concentration of protein \textit{y} decreases.
This generates a pulse in concentration of \textit{y} in response to a step input.
Such feedforward loops occur in diverse contexts in biology~\cite{yi00,goentoro09} and, recently, there have been synthetic demonstrations of such designs~\cite{guo2017}.
These examples present important work in identifying pulsed behaviour in biomolecular circuits, their functional roles, and underlying mechanisms.

There are at least three striking dynamical mechanisms that can underlie pulses.
First is excitability, a nonlinear dynamical phenomena with the signature that some small amplitude random inputs can trigger large changes in the output.
Second is based on a derivative-like effect on a step change in input implemented through integral feedback.
From a linear systems theory point of view, the circuit is said to have a zero at the origin ($s_0 = 0$, $s_0$ is a complex number) because a step change in input results in a net zero change in the value of the output at equilibrium.
More generally, a circuit has a zero at a generalized frequency $s = s_0$ if a change in input of magnitude $u = u_0e^{s_0t}$, where $t$ is time, results in a net zero change in the value of output at equilibrium~\cite{astrom2010feedback}.
When $s_0 = 0$, the change in input has magnitude $u_0$, which is a step change.
In case the integral feedback in the biomolecular circuit has an approximate, not exact, integral, the derivative is also approximate and the zero is close to the origin.
Third is non-normality, where all initial conditions decay to the equilibrium, but some of these initial conditions are such that the output grows transiently before an eventual decay.
Whether or not non-normality can facilitate
pulsing in biomolecular circuits has not been addressed in the
literature.

Here we ask if pulsing in feedforward loops is related to non-normality.
We addressed this using standard models of feedforward loops.
We found that non-normality can underlie pulsing in the standard incoherent feedforward loop.
We develop and apply a mathematical framework based on matrix norms and pseudospectrum analysis to identify non-normality in larger, more complicated circuits, finding that the presence of non-normality correlates well with pulsing.
We used these tools to obtain quantitative bounds on pulse shapes, to screen for pulse behaviour, and to design pulse dynamics.
These results provide a framework to understand and shape pulse dynamics in biomolecular circuits.

\section{Results}
\subsection{Non-Normality in Standard Incoherent Feedforward Loop}
We begin by investigating the role of non-normality in a standard model of an incoherent feedforward loop (Fig. 1b,~\cite{mangan03}).

\subsubsection{Model of an Incoherent Feedforward Loop}
In this model, a transcription factor $u$ acts as the input to the system, activating the expression of proteins $x$ and $y$.
Further, the protein $x$ is a transcriptional repressor of $y$.
With a step change in input $u$, the levels of proteins $x$ and $y$ increase.
As concentration levels of $x$ increase, they repress the production of protein $y$, and its concentration levels decline.
Therefore, a step change in input $u$ generates a pulse in the output $y$.

A simple mathematical model of this is ~\cite{shoval10},
\begin{equation}
\begin{split}
\frac{dx}{dt}&=\beta_xu-\alpha_xx,\\
\frac{dy}{dt}&=\beta_yu\frac{K_{xy}}{x}-\alpha_yy,
\end{split}\label{sys1}
\end{equation}
where $x$ and $y$ represents the concentrations of the proteins, $u$ is the input, $\alpha_x$ is the degradation rate for protein $x$, $\alpha_y$ is the degradation rate for protein $y$, $\beta_x$ is the production rate of protein $x$, $\beta_y$ is the production rate of protein $y$, and $K_{xy}$ is the dissociation constant for the binding of $x$ to the promoter of $y$.
Similar models have been used in the literature to study fold-change detection, scale invariance and perfect adaptation in feedforward loops~\cite{goentoro09,shoval11,sontag17,shoval10}. These models help to understand and quantitate the dynamics of this system. 

We numerically simulated Eqn. (1) for a step increase in input $u$ ($0.1\rightarrow 0.2$). The nominal values of other parameters were set to unity. As expected, we obtained a pulsed response of output $y$ (Fig. 2a, solid line).

\begin{figure}[!t]
\centering
\includegraphics[width=3in]{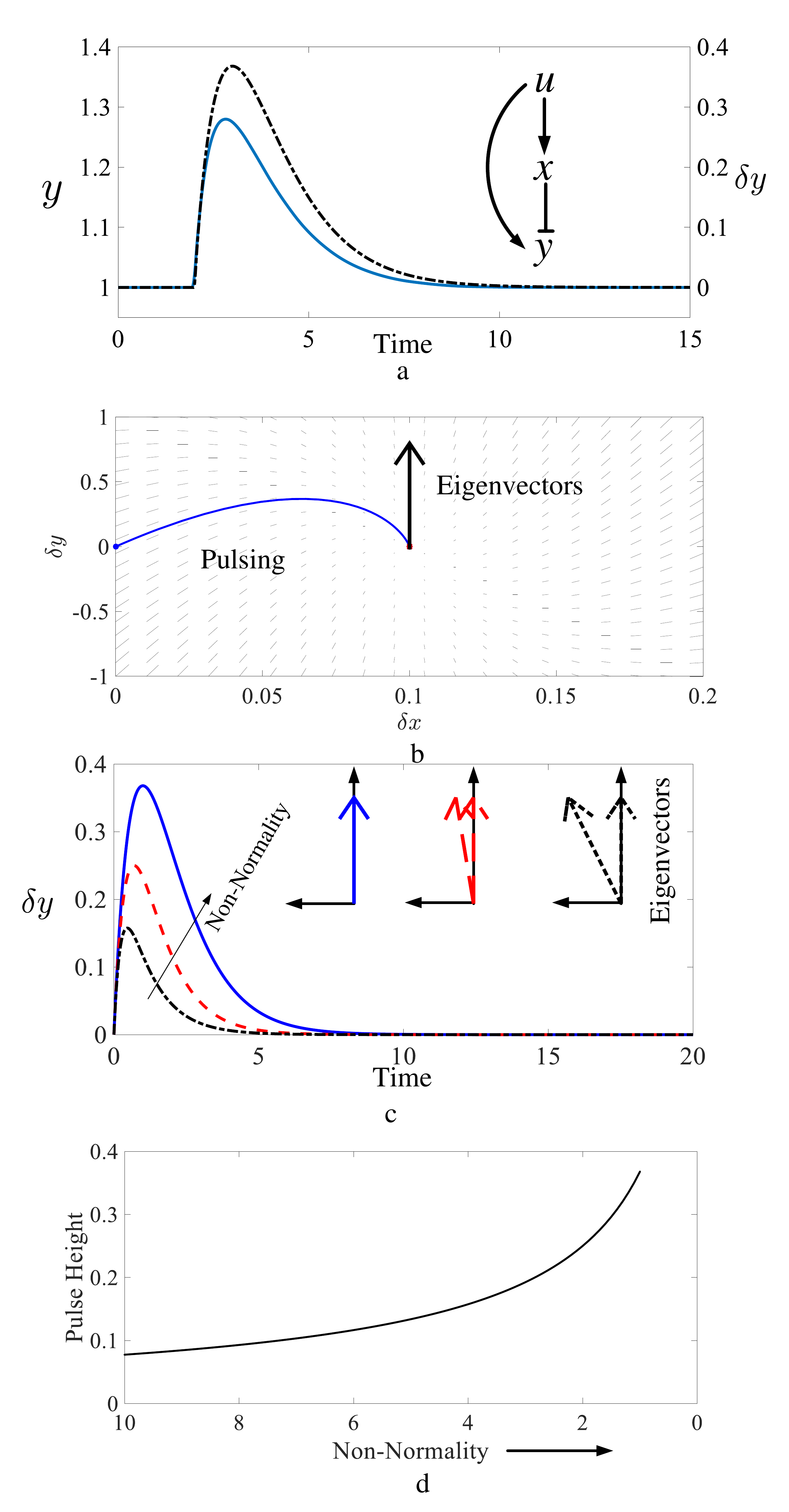}
\caption{a) Black solid line represents pulsed output for the full model in Eqn. (1) for a step change in input.
Black dashed line is the pulsed output for the linearized model in Eqn. (2). b) Phase plane for the full model. Blue line starting from the blue dot is the trajectory corresponding to the black solid line in a.
c) Blue solid line, red dashed line, and black dot-dash line are the pulse trajectories for the linearized model for different parameters.
These are for the parameters $\alpha_y$ =1, 2, and 4, respectively.
Remaining parameters are unchanged.
Corresponding eigenvector portraits are shown in inset.
d) Black solid line represents the pulse height as the extent of non-normality is changed through the parameter $\alpha_y.$}
\label{fig:1}
\end{figure}

\subsubsection{Investigation of Non-Normality}

To understand the modes of the system, we considered a linearization of Eqn. (1) at the pre-step equilibrium value (denoted $x_e$, $y_e$, $u_e$),
\begin{equation}
\delta\dot{z}=A\delta z+B\delta u,
\label{eq:2}
\end{equation}
where
\begin{equation*}
\delta z=\begin{bmatrix}
\delta x\\\delta y
\end{bmatrix},\ 
A\triangleq\begin{bmatrix}
-\alpha_x&0\\
-\frac{\beta_yK_{xy}\alpha_x^2}{\beta_x^2u_e}&-\alpha_y
\end{bmatrix},\ 
B\triangleq\begin{bmatrix}
\beta_x\\\frac{\beta_y\alpha_x K_{xy}}{\beta_xu_e}
\end{bmatrix}.
\end{equation*}
Similar to the nonlinear model, a step increase in $\delta u$ produces a pulse in the output $\delta y$ (Fig. \ref{fig:1}a, dashed line). For such a linear system, the natural solution is a linear combination,
\begin{equation}
\label{eq:state_solution}
\delta z(t)= c_1 v_1 e^{-\alpha_x t}+c_2 v_2 e^{-\alpha_y t},
\end{equation}
where $v_1$ and $v_2$ are eigenvectors of matrix $A$ corresponding to eigenvalues $\lambda_1$ and $\lambda_2$, and $c_1$, $c_2$ are constants depending on the initial conditions. For the dynamics represented in Eqn. (2), $\lambda_1=-\alpha_x$, $\lambda_2=-\alpha_y$, $v_1=[0\ 1]^T$, $v_2=[\frac{\alpha_y-\alpha_x}{H}\ 1]^T,$ where $H=-\frac{\beta_y\alpha_x^2}{\beta_x^2u_e}$. The angle between the eigenvectors, therefore, is determined by the system parameters. If the angle between the eigenvectors is $90^\circ$, the system is said to be normal. Non-normality, therefore, arises if the eigenvectors are not oriented orthogonally to each other.

The consequences of non-normality for the time evolution of trajectories can be seen as follows. The time evolution of the output in Eqn. (3) for a step change in input can be visualized with the help of eigenvectors. A step change in input changes the equilibrium point of the linearized system Eqn. (2). The path from the past equilibrium point to the new equilibrium point is determined by the eigenvectors (Fig. 2b). From Eqn. (3), the output $\delta z$ is obtained from the vector addition of the two vectors $c_1v_1e^{-\alpha_xt}$ and $c_2v_2e^{-\alpha_yt}$. If the system was normal and the eigenvectors were orthogonal to each other, this vector sum would decay monotonically to equilibrium. While the magnitude of each of these decreases with time, they may be oriented in such a manner that their vector sum transiently increases before decreasing. This is an indication of non-normality and can be analytically expressed as follows. We denote the magnitude of vectors $c_1v_1e^{-\alpha_xt}$ and $c_2v_2e^{-\alpha_yt}$ by $X$ and $Y$, respectively. The resultant can be obtained from parallelogram law as $R^2=X^2+Y^2+2XY \cos\theta$, where $\theta$ is the angle between the eigenvectors $v_1,\ v_2$~\cite{bale2010}. Note that when $\theta = 0^\circ$, the magnitude of $R$ is greater than when $\theta = 90^\circ$, $R$ may grow transiently before decaying to zero. The condition for this is,
\begin{equation}\label{eq:tr1}
  \left.\frac{dR}{dt}\right\vert_{t=0}>0.
\end{equation}

For the linearized model of the feedforward loop considered above (Eqn. (2)), there is transient growth if,\begin{equation}
\theta < \sin^{-1}\left(\frac{1-\bar{\alpha}}{1+\bar{\alpha}}\right),
\label{eq:2dcond}
\end{equation}where $\bar{\alpha}=\frac{\alpha_x}{\alpha_y}$ is the ratio of the eigenvalues. This condition can be used to understand the requirements for a pulse or to shape the pulse by modulating the parameters. We illustrate this by choosing different degradation rates $\alpha_x$ and $\alpha_y$ and correlating this choice with their effect on the pulse amplitude (Fig. 2c). When the eigenvectors are collinear, or approximately so, the largest growth is observed in the output pulse.
As the angle between eigenvectors increases, the pulse height starts to decrease. 
\subsubsection{Role of Zero}
The linearized equations of the model Eqn. (2) have a zero at $s = 0$.
This can be checked by setting the change in input $\delta u(t) = u_0e^{s_0t}$.
As this is a linear system, this implies that $\delta z = z_0 e^{s_0t}$.
For $s_0$ to be a zero, the change in output at equilibrium, $\delta y  = 0$, this implies $C\delta z = 0$.
From this, we get that for $s_0 = 0$, the output for a step change in input is zero.
Therefore, this is similar to a derivative-like mechanism to generate pulses in this model as well.

To check the interplay of the system zero and non-normality in shaping the pulse dynamics, we reconsidered parameters with different extents of non-normality.
We find that as the extent of non-normality increases, the value of the zero staying the same, the pulse amplitude increases.
This is represented in Fig. 2d.
These results show that, from a quantitative point of view, non-normality can be used to change pulse properties.

We conclude that non-normality is inherent in the standard model of a feedforward loop circuit, and together with the system zero can shape the pulse behaviour.

\subsection{Non-Normality in Larger Biomolecular Circuits}
Next, we analyze the presence of non-normal dynamics in larger biomolecular circuits. We start by developing and adapting mathematical methods that provide a generalized framework for this.

\subsubsection{Matrix Norm}

Pulsing dynamics are not necessarily confined to systems with two proteins as in the model considered in previous section. For a model with `$n$' proteins, there are `$n$' differential equations (with $n=2$ for Eqn (1)). The corresponding linearization, similar to Eqn (2) is,
\begin{equation}
\delta\dot{z} = A\delta z+B\delta u,
\end{equation}
where $\delta z$, and $B$ are $n\times 1$ vectors, $A$ is an $n\times n$ square matrix, and $\delta u$ is scalar input.
To adapt sufficient conditions for pulsing behaviour, similar to Eqn. (5), we focus on the homogeneous part of the equation containing the matrix $A$~\cite{schmid2007}. The solution of the equation $\delta\dot{z}=A\delta z$, involves the exponential of a matrix, $\delta z(t)=e^{At}\delta z(0)$, where $t=0$ is the initial time and $\delta z(0)$ is the initial condition. The growth of solution is tightly upper bounded by the matrix norm $R =\max_{\delta z(t_0)}\dfrac{\|\delta z(t)\|}{\|\delta z(t_0)\|}$, where $\|.\|$ is the length of vector $\delta z(t)$. The sufficient conditions for transient growth, similar to Eqn. (4), is $\left.\dfrac{dR}{dt}\right\vert_{t=0}>0$.

The maximum value of this can be explicitly calculated for stable matrices $A$, whose eigenvalues have a real part that is negative. This maximum value is $W(A) = \sigma_{max} (A + A^T)$, where $T$ is the transpose of $A$ and $\sigma_{max}(X)$ is the maximum eigenvalue (proof from~\cite{schmid2007} reproduced in S1). If $W(A)$ is positive it means this is a sufficient condition for $\left.\dfrac{dR}{dt}\right\vert_{t=0}>0$ and there is an output pulse. For model considered in previous section, $W(A) = 8 \geq 0$, consistent with a pulse in the output.

\begin{figure}[!htbp]
\centering
\subfigure[]
 {
  \includegraphics[width=3.5in]{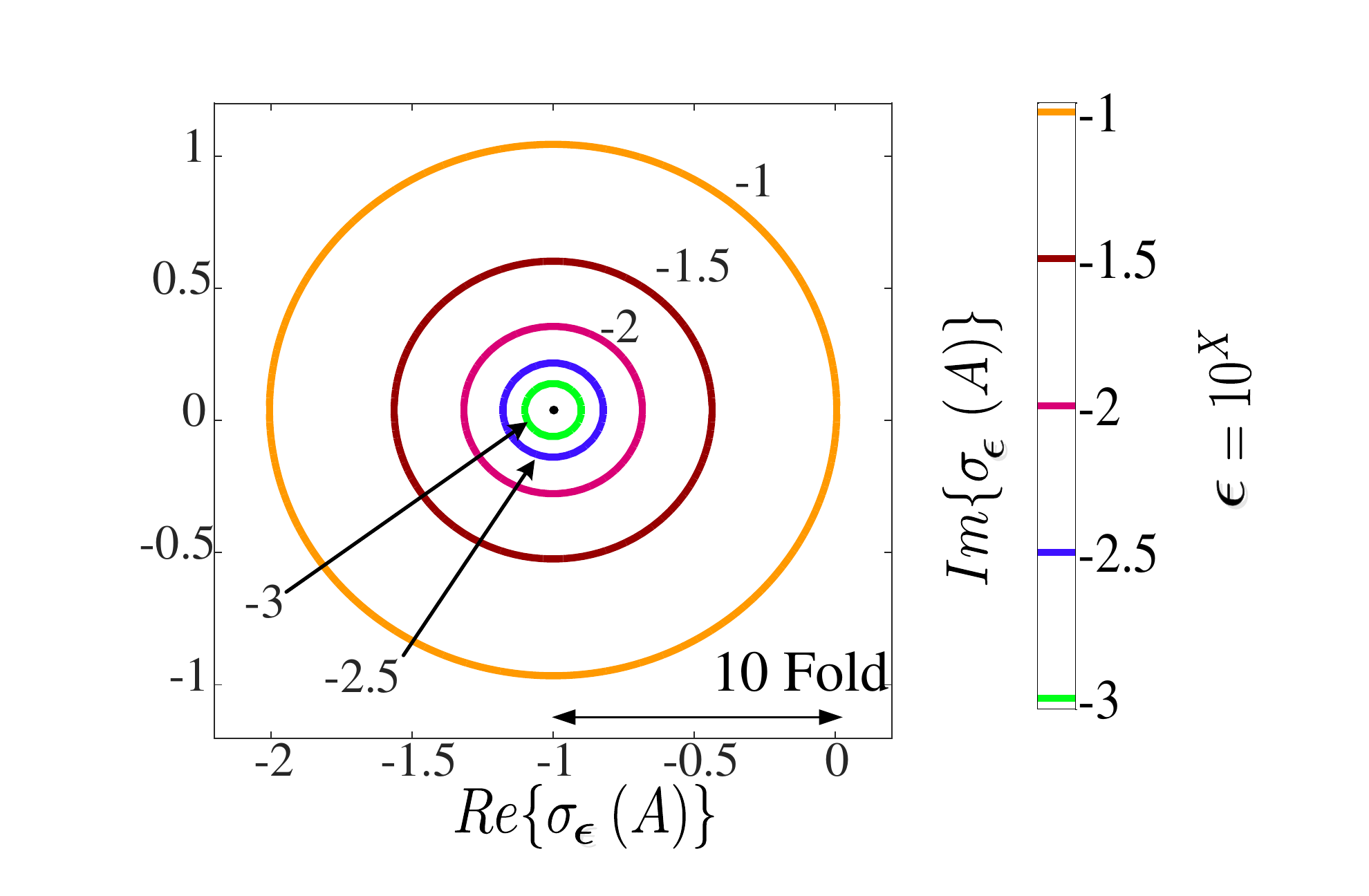}
  \label{}
  }
 \subfigure[]
 {
  \includegraphics[width=3.5in]{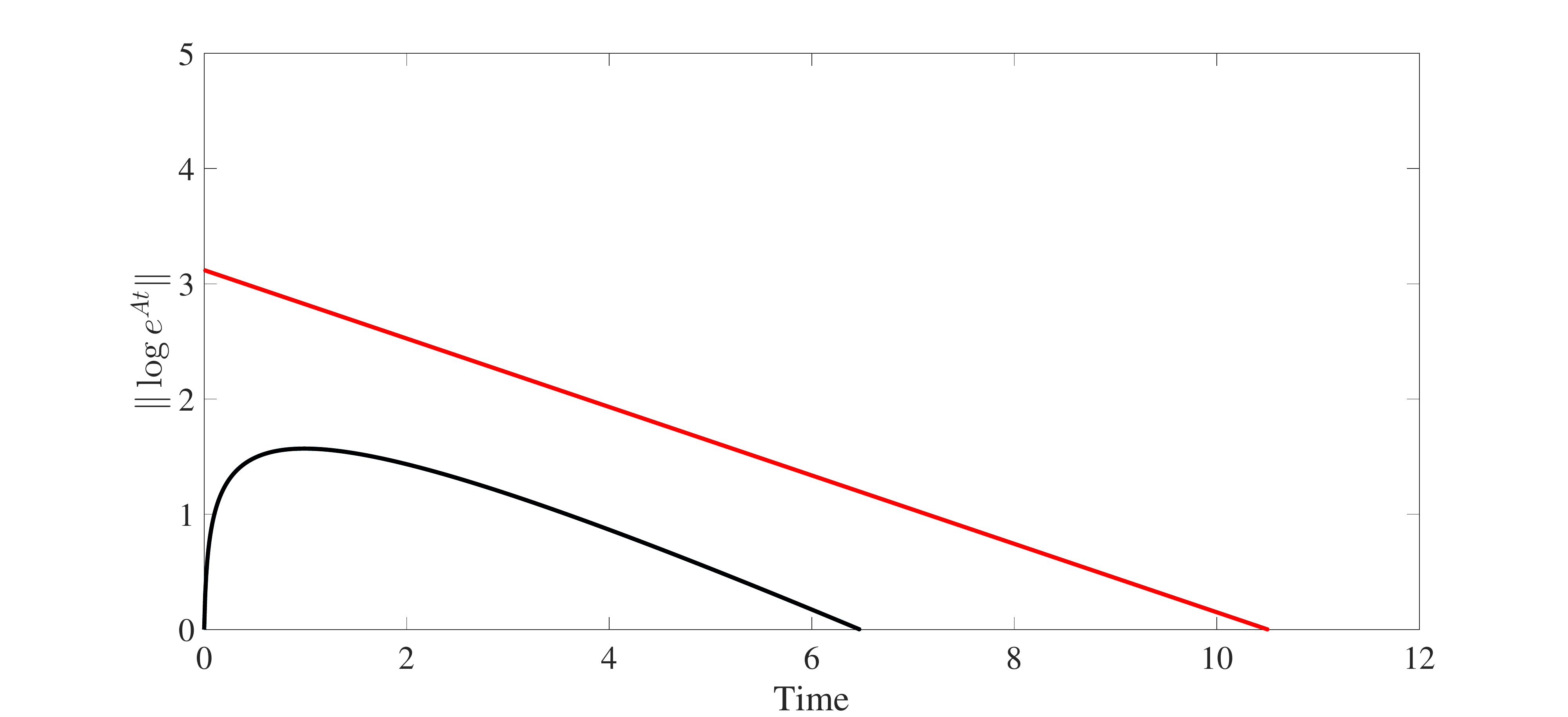}
  \label{}
  }
\caption{Application of non-normality tools on the incoherent feedforward loop model. a) Pseudospectral plot for the model in Eqn. (2).
Circles of different colours indicate the changes in the eigenvalue from the nominal value.
These colours correspond to different extents of perturbation $\epsilon$ as indicated in the colour bar. b)  Black solid line represents $\|e^{At}\|$ for the model in Eqn. (2). Red line is the bound computed for $\epsilon =0.01$.}
\end{figure}

\subsubsection{Pseudospectrum Analysis}

For models with two proteins, non-normality can be illustrated geometrically using eigenvectors (Fig. 2b). As the models become larger, the dimensions of the eigenvectors also increase and their geometric visualization is relatively difficult. Pseudospectrum analysis provides a tool for this~\cite{golub2012book,friedland81}. Below we present a simplified treatment of these. A more technical treatment is in the Supplementary S1.

The pseudospectrum of a square matrix $A$, denoted $\sigma_{\epsilon}(A)$ is the set of numbers $s$ which are eigenvalues of a matrix $A + E$, where $E$ is also a square matrix with $\|E\| < \epsilon$, $\epsilon$ being a small positive number. A system (Eqn (8)) is normal if the magnitude of perturbation in eigenvalues are lower than the magnitude of perturbation in its $A$ matrix. For a stable system, the upper and lower bounds on $e^{At}$ can be obtained as
$\dfrac{\gamma_\epsilon(A)}{\epsilon}\leq\sup_{t\geq 0}\|e^{At}\|\leq\dfrac{L_{\epsilon}(A)}{2\pi\epsilon} e^{\gamma_\epsilon t}$, where $L_\epsilon$ is the contour of the pseudospectrum and $\gamma_\epsilon$ is the maximum of real parts of elements of $\sigma_\epsilon (A)$. At $t = 0$, $\delta z(t) = \delta z(0)$ and $\|e^{A0}\| = 1$. For non-normality, $\gamma_\epsilon/ \epsilon > 1$ $\implies\|e^{At}\|$ may be greater than $1$ and there is a transient pulse.

We performed this pseudospectrum analysis on the feedforward model considered previously.
For this, we used \textit{EigTool}, a MATLAB based package~\cite{Senn:2009}. We found that small perturbations can generate large changes in eigenvalues (Fig. 3a). This shows that a pulsed output is expected from a pseudospectrum analysis. An additional advantage of the pseudospectrum analysis is the bounds that can be obtained. For the feedforward loop model considered previously, these bounds agree well with that is observed numerically (Fig.3b)


\subsubsection{Non-Normality in 3-Node Circuits}
To test the role of non-normality in generating pulse dynamics in larger circuits, we used the above-mentioned methods. First, we considered a set of randomly generated three-node circuit topologies~\cite{skataric12}, some of which can exhibit a pulsed behaviour. We found that these circuits correlated well with inherent non-normal dynamics.

We illustrate this analysis using an example from \cite{skataric12}. The mathematical model of this circuit is,
\begin{equation}
\begin{split}
\dot{x}_1&=k_1u-k_2x_2,\\
\dot{x}_2&=k_3x_1-k_4x_2,\\
\dot{x}_3&=k_5x_1-k_6x_2x_3.\\
\end{split}
\end{equation}
where, $u$ is the input, $x_i$ ($i=1,\ 2,\ 3)$ are the circuit nodes and $k_j(j=1,\ 2\ ...,6)$ are the reaction rate constants.
For this model, the parameters were set to $k_j=1$ for $j\neq4,\ k_4=2$.
A step change in $u$ produces a pulse in $x_3$ (Fig. 4a).

The linearized model $\delta\dot{z} = A \delta z+B\delta u$ has three eigenvalues $\lambda_1=-0.1,\ \lambda_2=\lambda_3=-1$. The eigenvector corresponds to $\lambda_1=-0.1$ is $v_1=\begin{bmatrix}0&0&1\end{bmatrix}^T$, and eigenvector corresponding to $\lambda_2=\lambda_3$ is $v_{2}=\begin{bmatrix}0.55&0.55&0.62\end{bmatrix}^T$. As there are third order eigenvectors, a visual illustration of non-normality is harder relative to the simpler two-dimensional model analyzed in the previous section. Therefore, we computed the abscissa $W(A)$ and the pseudospectrum. For this circuit, $W(A) = 1.313$, which is positive. This implies that there is initial transient growth consistent with the observed pulsed output.
Then, we computed the pseudospectrum and find that a perturbation of $\epsilon=0.1$ gives 5-fold change in eigenvalue (denoted Fold$_\epsilon$ in Fig. 4a), implying the presence of non-normality.

We repeated this analysis for the entire set of topologies that exhibited pulsed dynamics in~\cite{skataric12}.
Our aim of investigating these is to check the role of non-normal dynamics in the pulsed behaviour.
We estimated the presence of non-normal dynamics using the tools of matrix norms and pseudospectra.
We found that all topologies that pulsed had inherent non-normal dynamics.
These results are tabulated in Supplementary Material S2.

\begin{figure}[htbp]
\centering
\includegraphics[width=3in]{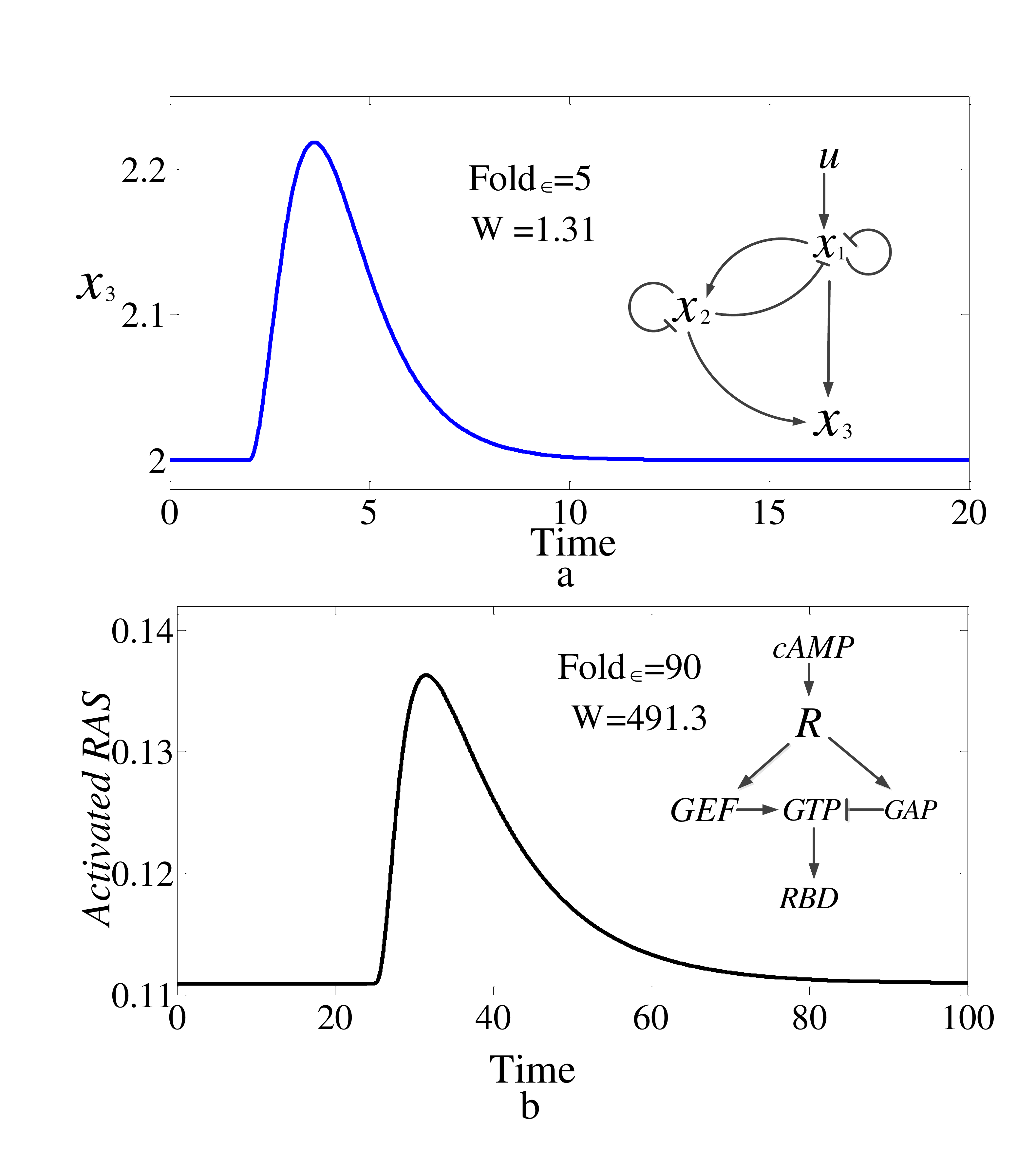}
   \caption{Non-normality in larger biomolecular circuits. a) Black solid line is the pulse output $x_3$ in the three node pulsing circuit (inset) for a step change in input ($u: 0.1\rightarrow 0.2$). b) Black solid line is the pulse output $Ras^{GTP}$ when $cAMP$ changes from $1$ to $2$ in adaptive $Ras-cAMP$ signaling network (inset).}
\end{figure}

\subsubsection{Non-Normality in Eukaryotic Chemotaxis Signaling Pathway}
Next, we analyzed the role of non-normality in the pulse behaviour of a eukaryotic signaling pathway~\cite{takeda12}.
This system adapts to changes in chemoeffector cyclic adenomonophosphate (cAMP).
A mathematical model of this system has been developed previously ~\cite{skataric12}.
It models two different affinity receptors \textit{$R_1$} and \textit{$R_2$} that bind \textit{cAMP}.
This complex activates two proteins \textit{Ras-GEF} and \textit{Ras-GAP} that incoherently act on activated \textit{Ras-GTP}.
Activated \textit{Ras} is measured using a cytosolic reporter \textit{RBD-GFP}.
The model equations are,
\begin{equation}
\begin{split}
\dot{x}_1&=k_1(u+r_1)(\bar{x}_1-x_1)-k_2x_1,\\
\dot{x}_2&=k_3(u+r_2)(\bar{x}_2-x_2)-k_4x_2,\\
\dot{x}_3&=k_5(x_1+x_2)-k_6x_3,\\
\dot{x}_4&=k_7(x_1+x_2)-k_8x_4,\\
\dot{x}_5&=k_9x_3(\bar{x}_5-x_5)-k_{10}x_4x_5,\\
\dot{x}_6&=k_{11}(\bar{x}_6-x_6)-k_{12}x_6x_5,\\
\end{split}
\end{equation}
where $x_1=R_1$, $x_2=R_2$, $x_3=GEF$, $x_4=GAP$, $x_5=Ras^{GTP}$, $x_6=RBD^{cyt}$ are the states of the system. The values of the system parameters are considered as in ~\cite{skataric12}. 

A step change in $cAMP$ results in a pulse of activated $Ras$ concentration (Fig. 4b).
To check if non-normality has a role in these dynamics, we computed the numerical radius $W$ and pseudospectrum of the linearized system obtained by linearization around the pre-step equilibrium point.
We find that the numerical radius $W = 491.3 > 0$ indicates the presence of the pulse.
We find that a perturbation of $\epsilon = 0.1$ gives $90$-fold change in eigenvalue, implying the presence of non-normality.

These results show that non-normality can underlie pulses in larger more realistic models of biomolecular circuits and highlights the utility of the numerical radius and pseudospectrum as tools that can be used to screen for pulsed behaviour.

\begin{figure*}[!tb]
\centering
\includegraphics[width=6in]{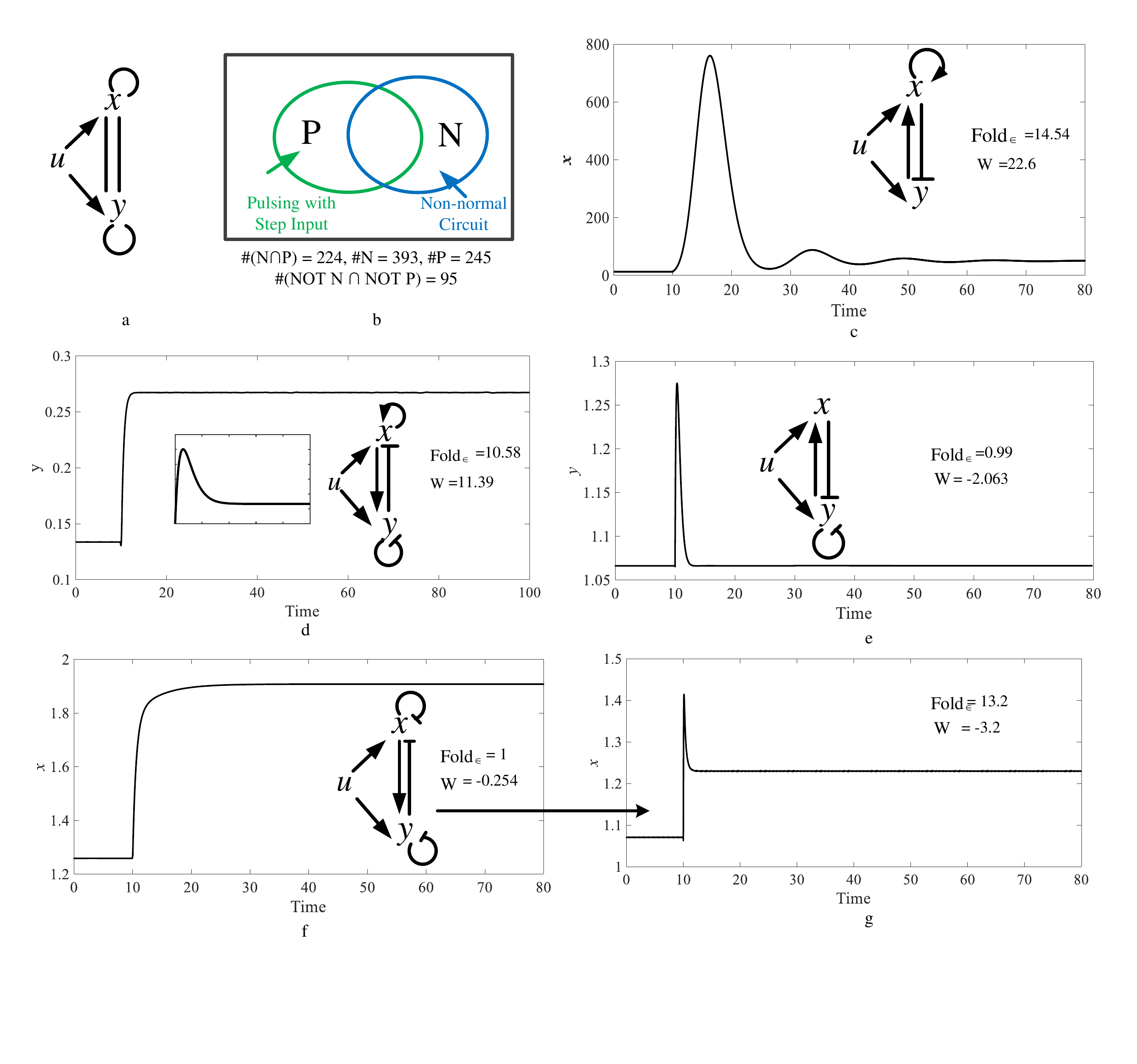}
   \caption{Screen of 2-node circuits using non-normality. a) Illustration of networks considered in the screen. The solid lines can be activating, inhibiting, or be absent. b) Venn diagram indicating the set of circuits that show a pulse, the set of circuits that are non-normal, and their intersection.  Specific examples from different parts of the Venn diagram showing the circuit, output response, and various non-normality metrics for c circuit that is non-normal and shows a pulse, d circuit that is non-normal does not show a pulse for a step input, e circuit that shows a pulse but is not non-normal as per the metrics used, f circuit that does not show a pulse and that is not non-normal as per metric used. Inset in d shows that circuit can exhibit a pulse from specific initial condition. g) Design of a circuit from e that is non-normal and exhibits a pulse in response to step input.}
\label{fig:2}
\end{figure*}
\subsection{Using Non-Normality for Screening Pulse Behaviour}
Given the role of non-normality and methods to determine non-normality in biomolecular circuits, it may be possible to use it for the purposes of screening pulse dynamics.
It may, in particular, add a layer of analytical support to data generated from numerical simulations.
To illustrate this, we constructed a library of 2-node circuits based on previous methods~\cite{ma09}.
Each of these circuit topologies had two nodes that were acted on by an input (Fig.5a).
Each node has the possibility of a positive, negative, or no interaction between itself and the other node.
Both nodes were viewed as outputs.
There are a total of $N = 3^4$ circuits.
For each circuit, we sample $M = 10$ parameter sets to obtain a total of $M\times N = 810$ circuits.
A typical model of a circuit is,
\begin{equation}\begin{split}
\dot{x}&=k_1uf_1(x,y)-k_2x,\\
\dot{y}&=k_2uf_2(x,y)-k_3y,
\end{split}\end{equation}
where $u$ is the input and $k_1$, $k_2$, $k_3$, $k_4$ are parameters of the system that take values in the range [0.01 2].
The functional forms of $f_1$ and $f_2$ determine the interaction between the two nodes.
We used approximated Hill functions to model activation $f_a(x) = \frac{x}{x+K} \approx \frac{x}{K}$ and repression $f_r(x)=\frac{K}{K+x}\approx \frac{K}{x}$, where $K$ is the dissociation constant.
In the case that both $x$ and $y$ act on a node, the resultant interaction term is a product of the individual terms.
So, if $x$ activates itself and is inhibited by $y$, $f_1(x,y) = \frac{x}{K_x} \times \frac{K_y}{y}$.

For each circuit, we allowed the circuit to settle to equilibrium and calculated the output for a step change in input u ($1\rightarrow 2$).
Out of a total of $810$ circuits, $509$ settled to an equilibrium.
These were categorized into circuits that exhibited a pulse and those that did not pulse.
The criteria for a pulse was that the peak of the output should be more than initial as well as final value of output. 
There were $245$ such circuits.

To use the tools developed above, we also linearized the model.
For the linearized model, we computed the metrics of non-normality such as eigenvectors, $W(A)$, and pseudospectrum as well as the system zeros.
These are tabulated in Supplementary Material S3.
There were $393$ circuits which were expected to pulse based on the non-normality metrics (either $W(A)>0$ or pseudospectral fold change $>$ 1).
These were compared against the full nonlinear simulations and this is illustrated using a Venn diagram (Fig. 5b).
We categorized the different sets of this diagram and present specific examples below.

The first set of circuits are those that are expected to pulse based on non-normality and exhibit a pulse. 
There are $224$ such circuits. 
An example of this is the circuit shown in Fig. 5c.
The second set of circuits are those that are expected to pulse based on non-normality but do not exhibit a pulse.
There are $169$ such circuits. 
An example of this is the circuit shown in Fig. 5d.
While this does not exhibit a pulse in response to a step input, it can exhibit a pulse in response to an initial condition perturbation (Fig. 5d inset).
The third set of circuits are those that are expected to not pulse based on non-normality but do exhibit a pulse.
There are only $21$ such circuits.
An example of this is the circuit shown in Fig. 5e.
As the estimation of non-normality is based on sufficient conditions, this is expected. We considered
each of this individually and found that these are indeed non-normal
(based on the eigenvectors, analysis highlighted in Supplementary
Material S3).
The fourth set of circuits are those that are expected to not pulse based on non-normality and do not exhibit a pulse. 
An example of this is the circuit shown in Fig. 5f.

\textit{Converting a normal system that does not pulse into a non-normal system which pulses:}
To explore a potential utility in designing pulsing behaviour in a circuit that does not pulse, we considered circuit in Fig. 5f. For this, we tuned the linearized system matrix A by perturbing
the parameters $k_i$ so that it becomes non-normal (Fig. 5g).
This demonstrates how the tools of non-normality can potentially be used to design pulses.

\section{Discussion}

Pulses are functionally important dynamical behaviour observed in biomolecular circuits, such as in feedforward loops.
Here, we investigated the dynamical mechanisms underlying such pulses and presented three main results.
First, we find that non-normality can play a role in generating and shaping pulses in standard models of a feedforward loop.
Second, we demonstrated how tools to assess non-normality, such as pseudospectrum analysis, can be used to check for pulsing dynamics in larger circuits, to provide quantitative bounds for pulse amplitudes,  to screen for pulsing behaviour, and to design pulse dynamics.
Third, we find that non-normal mechanisms can combine with system zeroes to generate pulses of larger amplitude.
These results provide a dynamical understanding for the generation of pulses in biomolecular circuits.

It is interesting to note how a pulse waveform is generated from a combination of two exponential waveforms, each of which is decaying.
In the standard model of the feedforward loop, the two decaying exponentials represent the dynamics along the direction of the eigenvectors.
The decay exponents are the eigenvalues.
In particular, it is how these decaying exponentials are combined, through a difference, that is the source of the pulse waveform.
At the initial and final times, the exponentials have a similar value.
Therefore, their difference is close to zero.
In the intermediatory period, there is a transient difference and this reflects in the pulse growth.

An important direction of future work is to use the analysis based on non-normality for design of pulse behaviour.
Starting from specifications of pulse properties, such a synthesis workflow should generate an appropriate linearization and then a biomolecular circuit realization.
Some analytical developments that may help with these is to extend these tools for the overall nonlinear mathematical model, possibly though the use of tools such as Lyapunov exponents.

Pulse dynamics are prevalent in multiple contexts in engineering and science.
Here, we have highlighted the role of non-normal dynamics in the pulsing mechanism prevalent in biomolecular feedforward loops.
This is a linear dynamical mechanism and so, as investigated here, can be directly used for screening for pulse behaviour in large scale models.
For the same reason, this can help in the design of biomolecular circuits with pulse dynamics.

%

\bibliography{Ref_PulsingIFFL_IET}
\bibliographystyle{IEEEtran}

\section*{Acknowledgements}
We thank anonymous reviewers for their valuable comments.
Research supported partially by Science and Engineering Research Board Grant SB/FTP/ETA-0152/2013. The first
author acknowledges the financial support from the DeitY
Fellowship (MI01233).

\end{document}